# Navigating Sparse Singlet Fission Chemical Space: An Intelligent Generative-Predictive Paradigm

*Longfei Lv[1,2], Li Fu[1,2*], Si Zhou[1,2], Lingzhi Zhao[3*], Jijun Zhao[1,2*]*

[1] Guangdong Basic Research Center of Excellence for Structure and Fundamental Interactions of Matter, Guangdong Provincial Key Laboratory of Quantum Engineering and Quantum Materials, School of Physics, South China Normal University, Guangzhou 510006, China

[2] MOE Key Laboratory of Environmental Theoretical Chemistry, South China Normal University, Guangzhou 510006, China.

[3] Guangdong Provincial Engineering Technology Research Center of Low Carbon and Advanced Energy Materials, School of Electronic Science and Engineering (School of Microelectronics), South China Normal University, Foshan, 528225, China

[*] Corresponding authors. Email: fuli@scnu.edu.cn (Li Fu); lzzhao@scnu.edu.cn (Lingzhi Zhao); zhaojj@scnu.edu.cn (Jijun Zhao)

**Abstract:** Singlet fission (SF) offers a promising route to surpass the Shockley-Queisser limit by converting a photoexcited singlet exciton into two triplet excitons, thereby enhancing photovoltaic energy conversion efficiency. However, efficient SF process requires stringent energetic requirements among low-lying excited states, which makes SF molecules intrinsically rare within the vast chemical space. This extreme sparsity poses a grand challenge for molecular discovery. Because of low hit rates and trial-and-error computational waste on nonviable structures, conventional high-throughput virtual screening faces significant constraints, even when accelerated by machine learning models. Here, we establish a synergistic generative-predictive framework for the targeted inverse design of SF molecules by integrating a structure generator, a properties predictor and a multi-criteria validation workflow. By continuously coupling generative exploration with SF predictive models, the framework progressively enriches SF species and achieves a success rate of approximately 90% in generating molecules that satisfy the target SF energetic criteria. High-throughput evaluation of about 100 million generated structures, with time-dependent density functional theory (TDDFT) validation on only a random 1% subset, confirmed a 90.8% success rate for SF candidates. Overall, we constructed an SF database of 283,559 candidates with favorable excited-state energetics and synthetic accessibility. From this database, we identified a key fragment strongly associated with SF requirements, namely, CN([O])N(C)[O]. These findings establish an efficient route to overcome the sparsity challenge in SF molecular discovery and provide interpretable design principles for the development of novel excited-state functional materials.

## 1. INTRODUCTION

Excited states of molecules govern the fundamental photophysical processes underlying modern optoelectronic technologies[1-5]. Specifically, the relative energetics of singlet and triplet excited states determine the performance of many excited-state processes, ranging from reverse intersystem crossing in thermally activated delayed fluorescence (TADF) to multiexciton dynamics in triplet-triplet annihilation (TTA)[6-8]. Among them, singlet fission (SF) has attracted particular attention because it converts one photoexcited singlet exciton into two triplet excitons[9-13], thereby mitigating thermalization losses and offering a promising strategy to surpass the Shockley-Queisser efficiency limit for photovoltaic energy conversion[14]. Efficient SF, however, requires a delicate energetic balance among the low-lying excited states, typically involving a favorable energy alignment of the lowest singlet excited state ($S_1$), lowest triplet excited state ($T_1$), and second-lowest triplet excited state ($T_2$)[15-18]. These stringent energetic requirements substantially narrow the pool of viable molecular candidates. As a consequence, experimentally validated SF materials remain largely confined to a few rigid molecular scaffolds, such as acenes and rylene derivatives, which often suffer from poor chemical stability and limited structural tunability[19-22].

To address these fundamental constraints, a central challenge is to efficiently discover new SF molecules that meet the stringent energetic requirements while remaining synthetically accessible within the vast and largely unexplored chemical space. To date, forward virtual screening remains the most widely adopted strategy for the computational discovery of SF molecules. By coupling *ab initio* calculations (typically based on the time-dependent density functional theory, TDDFT) with molecular databases, these "forward" approaches filter candidates that satisfy the requisite SF energetic criteria (i.e., $E_{S1} \geqslant 2E_{T1}$ and $E_{T2} \geqslant 2E_{T1}$). For example, Padula et al. carried out an extensive TDDFT screen of about 40,000 molecules, yielding roughly 200 promising SF candidates[23]. Similarly, Perkinson et al. computationally assessed 4,482 anthracene derivatives and identified 88 SF candidates[24]. Machine learning (ML) models have substantially accelerated excited-state property prediction and enabled screening campaigns on unprecedented scales[25-28]. Nevertheless, these advances largely adhere to the forward-discovery paradigm, confining the search to molecules sampled from existing chemical spaces. Indeed, in our recent study, screening more than 20 million molecules using a graph neural network (GNN) yielded only a few hundred promising SF candidates[29]. This severe efficiency barrier is not merely a consequence of model capability or computational cost, but rather reflects the intrinsic sparsity of SF molecules within the vast chemical space. Traditional "forward" screening is, by nature, a "trial-and-error" strategy that squanders immense computational resources on non-viable structures. This fundamental limitation necessitates a paradigm shift from database-dependent screening to inverse design strategies that can directly construct target-oriented SF molecules.

Generative ML models provide the ideal vehicle for this shift, capable of *de novo* creation of molecules with desired properties while bypassing the need for exhaustive screening. While various generative models—including genetic algorithm[30], reinforcement learning[31], and diffusion models[32]—have been used to explore unknown SF molecules, efficiently generating viable SF candidates remains challenging due to poor structural validity and limited synthetic accessibility, as well as the difficulty of satisfying energetic constraints across multiple excited states. In this work, we develop a collaborative generative-predictive framework to effectively address these challenges. By iteratively coupling a molecular generator based on long short-term memory (LSTM) with an excited-state properties predictor and a multi-criteria validation process, this framework continuously refines the structural validity and synthetic accessibility of the outputs. Consequently, it achieves an approximately 90% success rate in generating viable SF molecules and ultimately establishes a database of over 280,000 SF candidates at the TDDFT level. Crucially, transcending black-box generation, we identify a key structural motif favorable for SF via fragment-based decomposition. The mechanistic insights derived from this study not only validate our approach but also provide transferable design principles for a broader class of excited-state functional materials.

## 2. METHODS

Inspired by the complementary roles of creative exploration and analytical evaluation in human cognition, we developed an intelligent generative-predictive strategy to discover SF molecules, integrating molecule generation, property prediction, and multi-criteria validation (Figure 1). The generative model explored the chemical space by producing novel molecular structures with SF-characteristic structural patterns. Concurrently, the predictive model evaluated the generated molecules by estimating their excited-state energies, with particular emphasis on the energetic requirements relevant to SF. Candidates passing the initial predictive filter were then validated through TDDFT calculations and iteratively incorporated into the dataset to train the generative and predictive models. When the synergistic generative-predictive workflow converged, the resulting molecular generator navigated the sparse SF chemical space and proposed a large number of targeted candidate molecules, followed by multi-criteria virtual screening to construct the SF database. Finally, fragment-based analyses were performed on the SF database to extract intuitive structure-property rules for rational SF molecular design. Further details of this paradigm were described in the following subsections.

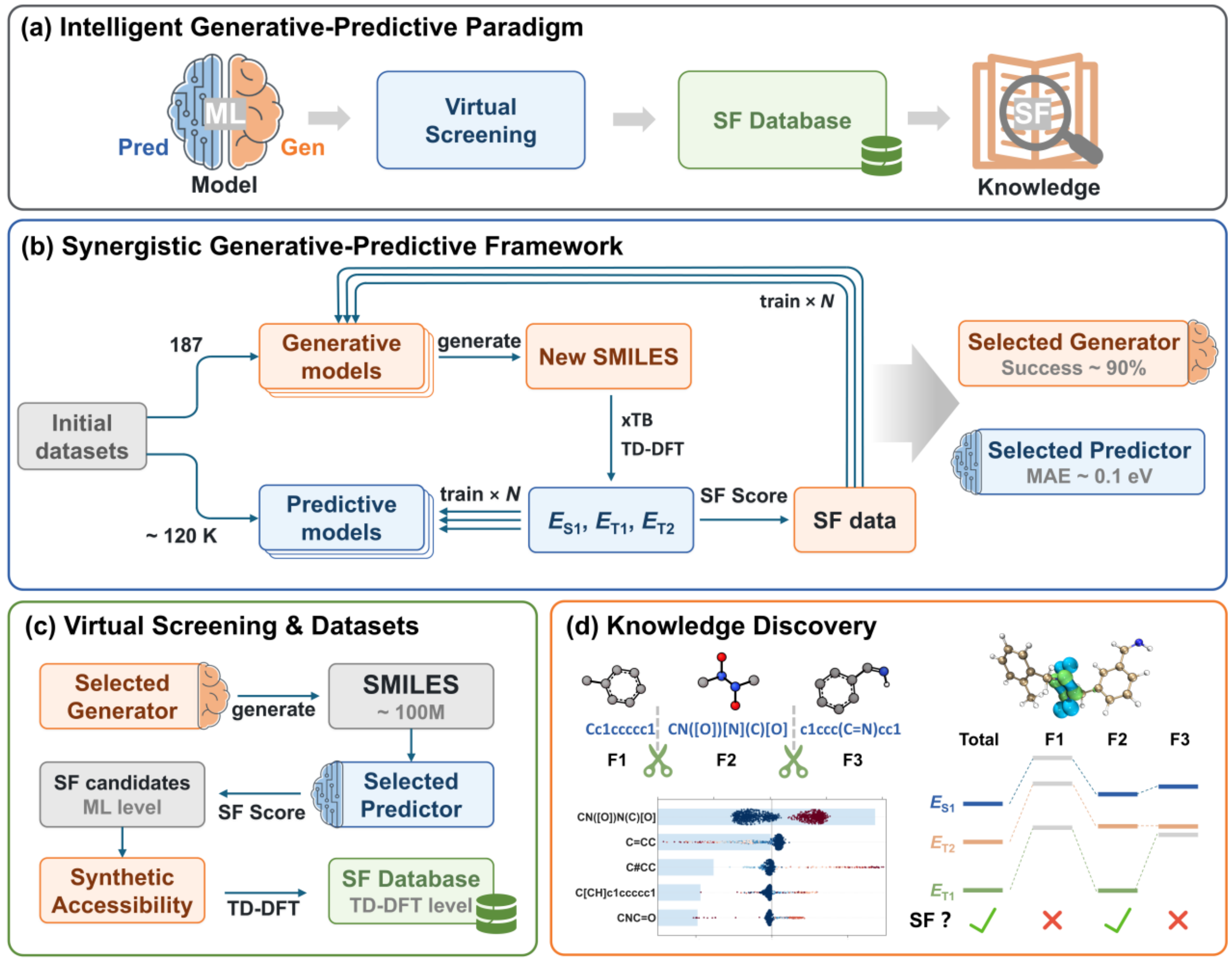

**Figure 1.** Overview of paradigm for SF molecular discovery in this work. (a) Intelligent Generative (Gen)-Predictive (Pred) Paradigm, which integrates Gen and Pred models to construct the SF database through virtual screening for knowledge discovery. (b) Iterative workflow of the synergistic generative-predictive framework. (c) Virtual screening with multi-criteria validation for SF database construction. (d) Knowledge discovery for elucidating the structure-property relationships governing SF performance.

### 2.1 Initial Datasets

Within the generative-predictive framework shown in Figure 1b, the initial dataset consisted of two complementary subsets. The first subset containing 187 SF molecules was screened by ML models and further evaluated by TDDFT calculations with the ωB97X-D functional[33] and 6-31G(d) basis sets[34], as reported in our recent publication[29]. These molecules served as the training set for the initial generative model, enabling the molecular generator to capture SF-relevant structural motifs. This dataset spans molecular sizes ranging from a few atoms to 50 atoms and covers 11 elements (C, H, N, O, F, Si, P, S, Cl, As, Se and Br). Given the limited number of available SF molecules, data augmentation using simplified molecular input line entry system (SMILES) representations was employed to expand the training set while preserving molecular identities (detailed in Supporting Information, Section S1). The augmented SMILES strings were then tokenized using chemistry-aware rules and zero-padded to a fixed sequence length for efficient model training[35, 36].

The second subset was taken from the FORMED database [37], which comprises 116,687 molecules constructed from experimental crystal structures with excited-state properties from TDDFT calculations. In brief, FORMED provides a broad coverage of chemical space, encompassing diverse molecular scaffolds, wide size range (4–260 atoms), and rich elemental compositions (C, H, N, O, F, P, S, B, Si, Cl, As, Se, and Br). This large-scale and chemically diverse dataset was used to train the initial property prediction model for learning generalized structure–property relationships.

### 2.2 Generative and Predictive Models

Both generative and predictive models operated on tokenized SMILES strings, shared a unified preprocessing pipeline, and employed a similar LSTM architecture to support two complementary tasks (Figure 2). An embedding layer first transformed discrete token indices into continuous low-dimensional vectors, which were subsequently processed by the LSTM or bidirectional LSTM (BiLSTM) network.

The generative model directly learned the probability distribution of valid SMILES sequences from the tokenized representations. The embedded token sequences were processed by LSTM layers, which updated hidden and cell states to capture sequential molecular information. A decoder mapped hidden representations to the vocabulary space, and the next token was sampled from the *Softmax* normalized probability distribution in an autoregressive manner until a full SMILES was produced. During training, the model was optimized to predict the next token in a sequence using cross-entropy loss and the Adam optimizer. A ReduceLROnPlateau scheduler was employed to decrease the learning rate when the validation loss ceased to improve, enabling more refined parameter updates during the later stages of training.

To screen SF candidates, the predictive model was designed to predict $E_{S1}$, $E_{T1}$ and $E_{T2}$ directly from input sequence embeddings based on SMILES. Embedded token vectors were processed by a BiLSTM, enabling each token representation to incorporate contextual information from both preceding and subsequent positions in the sequence. The global sequence features and final hidden states were then fed into an enhanced feature extraction block. In this block, a gated linear unit filtered informative features, and residual connections combined with layer normalization ensured stable information propagation. Finally, a multilayer perceptron regressor mapped the high-dimensional features into scalar excitation energies. The predictive models were trained by minimizing mean squared error between calculated and predicted excited-state energies using Adam optimizer. Adaptive learning-rate scheduling and gradient clipping were further employed to improve training stability.

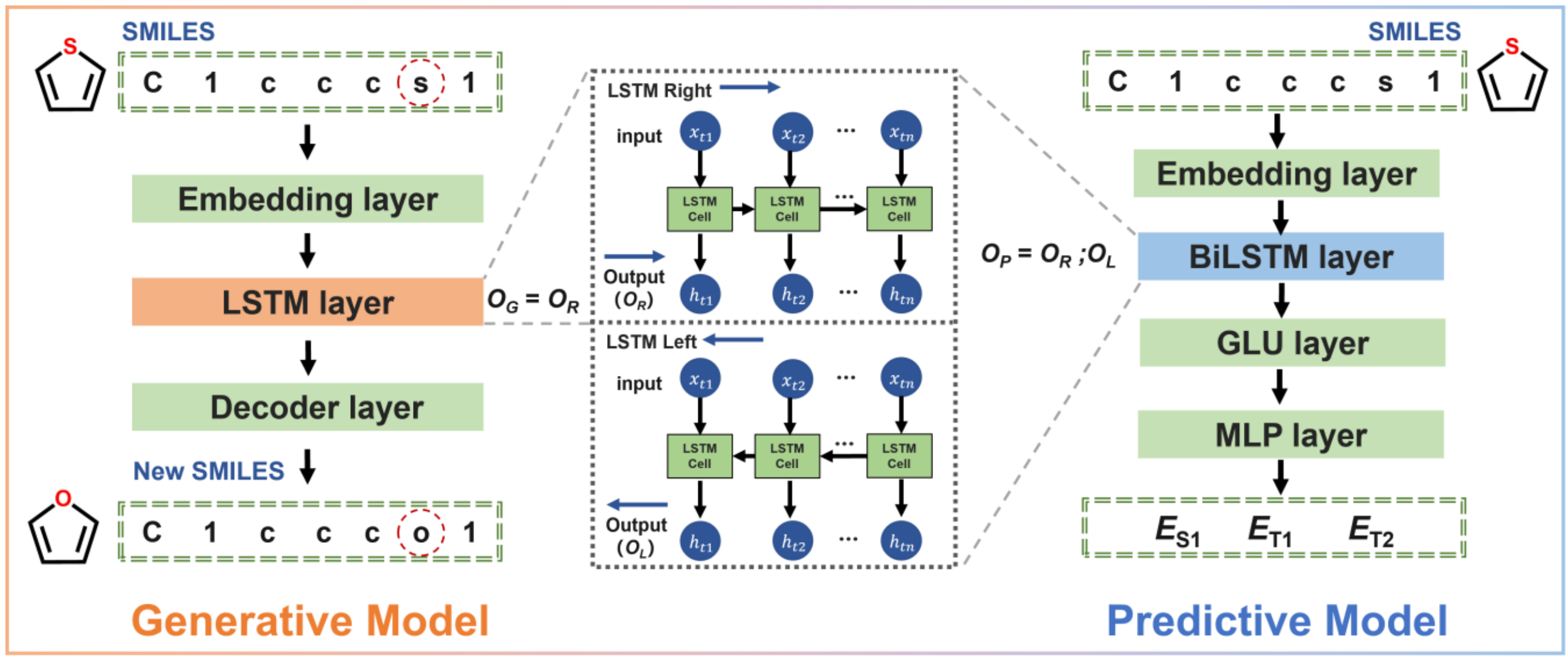

**Figure 2.** Schematic diagram of the structures of generative and predictive models based on SMILES and LSTM architectures.

### 2.3 Structure Optimization and TDDFT Calculation

The initial three-dimensional geometric structures were generated from the SMILES of molecules using the MMFF94 force field[38], which had demonstrated reliable performance in generating chemically reasonable molecular conformations[39, 40]. The MMFF94-generated molecular geometries were then relaxed using a semiempirical GFN2-xTB method[41, 42], which offers a reasonable balance between computational efficiency and structural accuracy for high-throughput calculations of organic molecules [43-45]. The reliability of GFN2-xTB-optimized geometries for subsequent excited-state properties had been demonstrated by Corminboeuf et al. through a benchmark study on 4000 structurally diverse molecules[37]. Their results showed that GFN2-xTB and ωB97X-D/6-31G(d)[33, 34] optimizations yielded highly consistent geometries, and that the corresponding $E_{\mathrm{S1}}$ and $E_{\mathrm{T1}}$ values coincided excellently. In this work, the satisfactory consistency between GFN2-xTB and ωB97X-D/6-31G(d) optimized geometries was further demonstrated (Section S2 in Supporting Information). Vertical excitation energies were obtained using Tamm–Dancoff approximated TDDFT (TDA-TDDFT)[46] at the ωB97X-D/6-31G(d) level of theory, as implemented in the Gaussian 09 suite[47].

### 2.4 Score of Singlet Fission

In general, a potential SF molecule has to satisfy several energetic criteria, namely, $E_{\mathrm{S1}} \geq 2E_{\mathrm{T1}}$, $E_{\mathrm{T2}} \geq 2E_{\mathrm{T1}}$ and $E_{\mathrm{T1}} \geq E_{\mathrm{g}}$ (where $E_{\mathrm{g}} = 1.1$ eV for silicon crystal)[29], as illustrated in Section S3 in Supporting Information. The first criterion ensures that the SF process is thermodynamically favorable, whereas the second one suppresses TTA by maintaining a sufficient energy separation between the first and higher-lying triplet states. The third criterion requires the triplet energy to exceed the band gap of the interfaced photovoltaic material, enabling efficient utilization of generated triplet excitons. To integrate these energetic requirements into a unified metric, we defined a continuous SF Score that simultaneously evaluated all three criteria as follows:

$$\xi = \begin{cases} \sqrt{c_1 \cdot c_2} & \text{if } c_1, c_2, c_3 \geq 0 \\ c_3 & \text{if } c_3 < 0 \\ min[c_1, c_2] & \text{else} \end{cases} \quad (1)$$

where $c_1 = E_{\mathrm{S1}} - 2E_{\mathrm{T1}}$, $c_2 = E_{\mathrm{T2}} - 2E_{\mathrm{T1}}$, $c_3 = E_{\mathrm{T1}} - E_g$. A molecule with positive SF Score is considered an SF candidates, whereas violation of one or more criteria results in negative score whose magnitude reflects the degree of energetic mismatch.

## 3. RESULTS AND DISCUSSION

### 3.1 Synergistic Generative-Predictive Framework

To overcome the inherent constraints of limited dataset size and insufficient coverage of chemical space, an iterative optimization mechanism was employed to guide the generation of molecular candidates, as illustrated in Figure 1b. This workflow was used to train a series of generative and predictive models, from which a generator and a predictor were selected for subsequent molecular exploration. The initial generative and predictive models were trained on the primary datasets described in Section 2.1. Dataset partitioning for the generative model followed a random 8:1:1 split into training, validation, and test subsets. Conversely, to ensure consistent target-value distributions across subsets, the predictive dataset was stratified according to property values before applying the same 8:1:1 split. Grid search optimization was independently performed to fine-tune the hyperparameters for both the initial generative model and the predictive models for $E_{S1}$, $E_{T1}$ and $E_{T2}$ (Table S1), and the resulting hyperparameter settings were retained throughout all subsequent iterations.

To evaluate the performance of the molecular generator, ~20,000 new SMILES strings for SF molecules were first generated using the initial trained generative model (Gen1). Following 3D structural generation and geometric optimization with the GFN2-xTB method, TDDFT calculations were performed to screen the candidate pool based on the SF Score. As shown in Figure 3a, the initial success rate of Gen1 model was only 20.8%. Here, the success rate for SF is defined as the proportion of candidate molecules with positive $\xi$ (4,665 in Gen1) among all molecules generated by the generative model (22,396 in Gen1) and validated by TDDFT calculations. These 4,665 newly identified SF SMILES strings were subsequently incorporated into the training set for the next-generation model (Gen 2). This iterative process was repeated sequentially for subsequent generations (Gen 3, Gen 4, etc.) until the performance of model converged. With an increasing number of SF SMILES across successive iterations, canonical SMILES and a deduplication filter were used to ensure both unique representation and novelty, yielding only unrecorded candidates.

As the iteration proceeded, the success rate for generating SF candidate molecules gradually increased and converged to approximately 90% at the sixth iteration (Gen 6). The insets in Figure 3a compare the distributions of $E_{S1}$, $E_{T1}$, and $E_{T2}$ for the generated molecules at iterations 1 and 6. By iteration 6, the generated molecules are predominantly clustered in the regions associated with positive $\xi$, indicating that the collaborative generative-predictive framework progressively steers molecular generation toward structures that satisfy the SF screening criteria. *t*-distributed stochastic neighbor embedding (*t*-SNE) analysis further reveals that the initial database only occupies a limited region of chemical space. In contrast, the molecules generated over six iterations extend well beyond the domain of the original dataset and populate the previously unexplored regions of chemical space. In comparison, a Transformer-based generative model was also evaluated. The current LSTM-based model outperformed it by generating more valid SMILES per iteration (see Table S2), which reflects its greater suitability for relatively small training datasets.

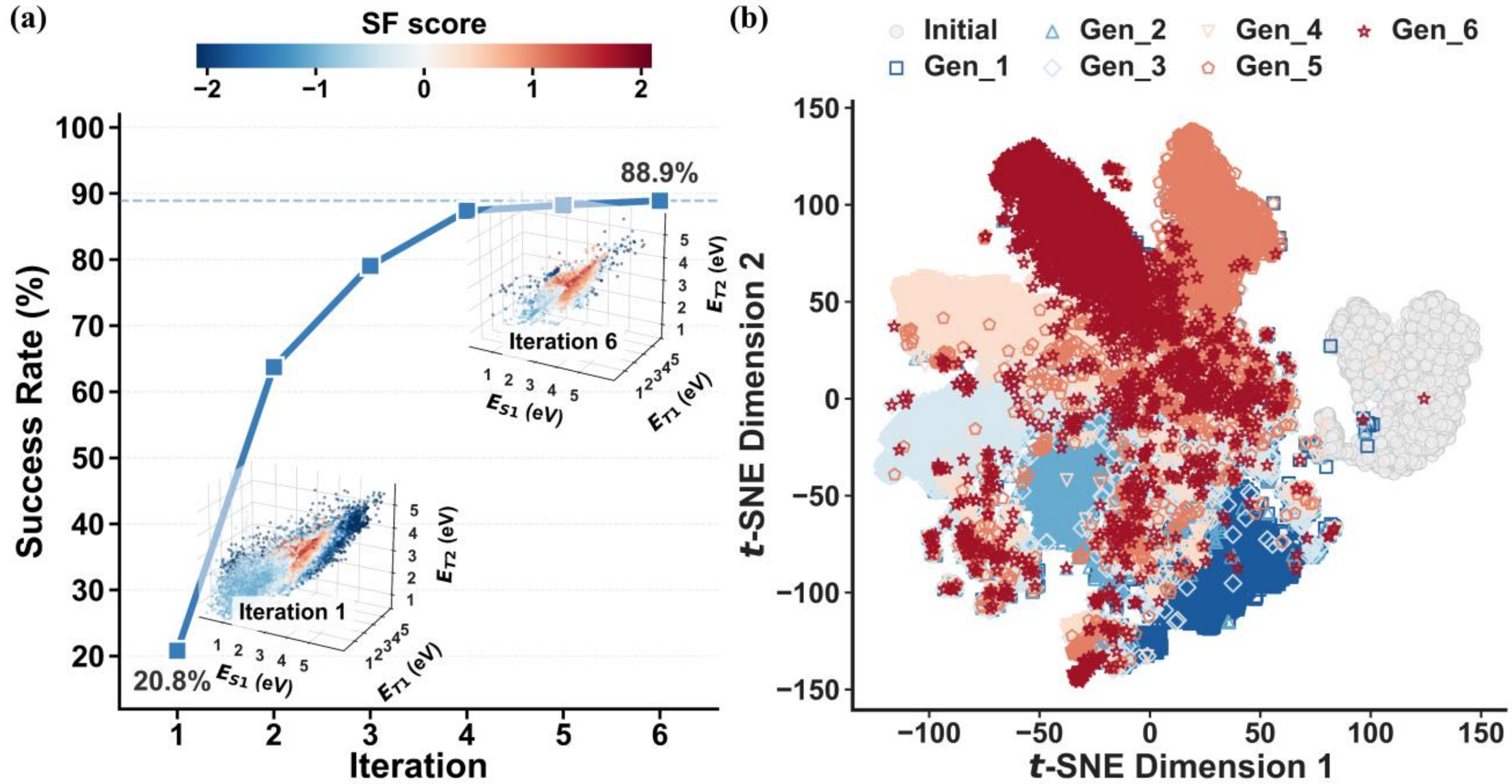

**Figure 3.** (a) Evolution of the SF success rate over six iterative generation cycles. Insets show the distributions of calculated molecules in the $E_{S1}$, $E_{T1}$ and $E_{T2}$ energy space at iterations 1 and 6, colored by their SF Scores. (b) *t*-SNE projection of the chemical space of SF molecules generated by different generative models. For each model, 4000 molecules were randomly sampled.

**Table 1.** Comparison of the MAE of LSTM (Pred1), SchNet, GNN, and XGBoost models trained on the FORMED dataset for predicting the HOMO-LUMO gap ($E_{H-L}$), $E_{S1}$, and $E_{T1}$.

| Models | MAE of $E_{H-L}$ (eV) | MAE of $E_{S1}$ (eV) | MAE of $E_{T1}$ (eV) |
|---|---|---|---|
| XGBoost[37] | 0.26 | 0.20 | 0.18 |
| SchNet[48] | 0.30 | 0.20 | 0.17 |
| GNN[29] | 0.16 | 0.11 | 0.08 |
| LSTM (this work) | 0.22 | 0.15 | 0.13 |

On the FORMED dataset, the first predictive model (Pred1) achieved mean absolute error (MAE) of 0.15, 0.13, and 0.12 eV for $E_{S1}$, $E_{T1}$ and $E_{T2}$, respectively (Figure S3), and outperformed both XGBoost and SchNet models (see Table 1). Although the GNN model from our previous work offers slightly higher accuracy, its dependence on accurate 3D structures (requiring geometry optimization) substantially raises computational cost and hinders its use in high-throughput screening of new SMILES. As the generative-predictive framework iteratively proceeded, the size of the TDDFT dataset for excited-state properties increased, as listed in Table S3. These newly acquired data were added to the training dataset for the next iteration of the predictive model, thereby progressively improving the predictive capability. After four iterations, the prediction errors for newly generated molecules converged to approximately 0.1 eV (Figure 4a; additional details are provided in Section S4), i.e., MAEs of 0.08, 0.06, and 0.07 eV for $E_{S1}$, $E_{T1}$, and $E_{T2}$, respectively. At successive iterations, the performance of Pred5 and Pred6 models shows no further improvement. The parity plots in Figure 4b–d also demonstrate satisfactory agreement between TDDFT-calculated and Pred4-predicted energies, with most data points clustered near the $y = x$ line.

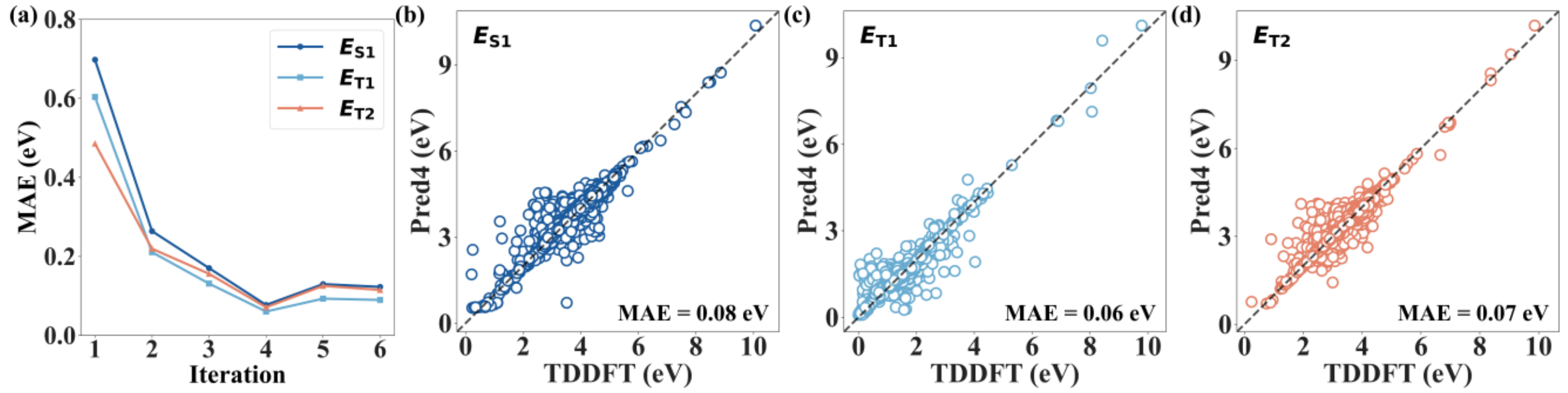

**Figure 4.** (a) MAE of $E_{S1}$, $E_{T1}$ and $E_{T2}$ for the TDDFT-validated generated molecules during the iterative process of the prediction model. Scatter plots of Pred4 model vs. TDDFT calculations for the generated molecules: (b) $E_{S1}$, (c) $E_{T1}$, and (d) $E_{T2}$.

### 3.2 High-throughput Screening and Database of SF Candidates

As schematically shown in Figure 1c and 5a, high-throughput screening of SF candidates was conducted by integrating generative-predictive models with multiple filtering criteria. Specifically, the Gen6 model was employed to construct an initial SMILES library of approximately 100 million molecules, and the Pred4 model was employed to predict their $E_{S1}$, $E_{T1}$, and $E_{T2}$ values. This ML-based screening retained nearly 85 million candidates with positive SF Scores, a yield consistent with the success rate achieved in the converged generative-predictive workflow. The remaining candidates were then evaluated for synthetic accessibility using the DeepSA model[49]; applying a threshold of 0.5 (above which a molecule is considered easy to synthesize) identified approximately 28 million candidates possessing both promising SF characteristics and high synthetic feasibility. Given the vast number of candidates, a representative subset of 1% (275,582 candidates) was randomly selected for TDDFT validation, which confirmed 250,138 molecules satisfying the SF energetic criteria (corresponding to a remarkably high success rate of 90.8%). On the other hand, the iterative generative-predictive training workflow in Section 3.1 yielded 86,128 SF candidates, among which 33,854 molecules have DeepSA scores exceeding 0.5. Adding these two datasets together and removing duplicate SMILES, we obtained a final TDDFT-calculated database comprising 283,559 unique and "easy-to-synthesize" SF molecules. It catalogs comprehensive structures and excited-state properties—including $S_1$-$S_5$ and $T_1$-$T_5$ excitation energies, oscillator strengths, and SF Scores—calculated at the TDDFT level (detailed in Supporting Information, Sections S5 and S6).

Pareto front analysis was employed to evaluate the simultaneous optimization of SF properties and synthetic accessibility within the SF database (Figure 5b)[50]. A significant population of molecules clusters in the high-density region where the DeepSA score approaches 1.0, demonstrating that the virtual screening strategy inherently prioritizes candidates possessing chemically reasonable structures and high synthetic accessibility. Structural classification revealed a vast and diverse pool of candidates. As illustrated in Figure 5c, six representative structural classes are highlighted: acene, amide, imide, hydrazone, nitrile, and quinoxaline[31]. These classes encompass considerable diversity in conjugated backbones, heterocyclic motifs, and functional groups. Notably, beyond these conventional families, the SF database encompasses a wealth of unconventional molecules that defy simple taxonomic categorization. This extensive structural diversity, coupled with the presence of unprecedented frameworks, confirms that our generative model effectively explores uncharted chemical space rather than merely iterating on established SF chromophores, thereby providing innovative candidates for experimental validation.

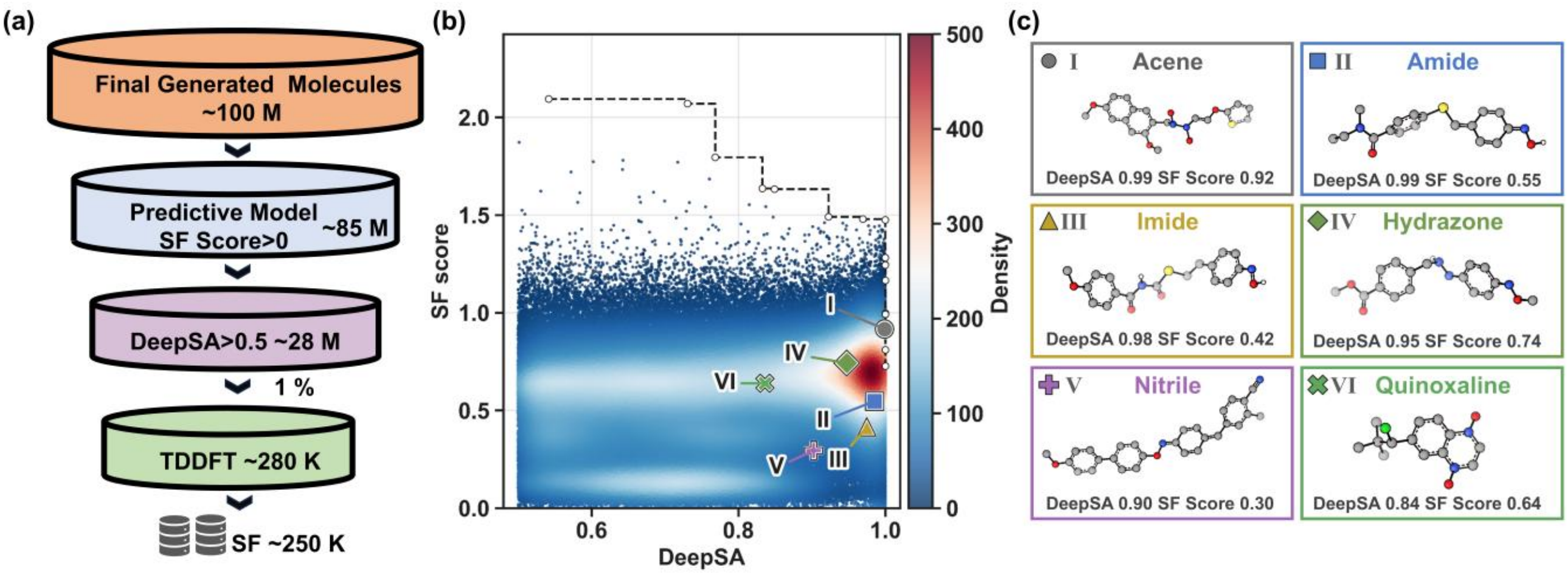

**Figure 5.** Multi-criteria high-throughput screening workflow for SF molecules and property distribution analysis. (a) Multi-criteria screening of generative molecular toward a TDDFT-validated SF database. (b) Density scatter plot of SF Score versus DeepSA score for screened candidates, highlighted with six molecular scaffolds. (c) Molecular structures, DeepSA and SF Score values of six scaffold families.

### 3.3 Knowledge Discovery

To identify the structural determinants governing SF properties, we randomly selected two balanced subsets (50,000 molecules each, with $\xi \geq 0$ and $\xi < 0$, respectively) from the TDDFT-validated dataset. These molecules were represented using a custom 526-bit molecular fingerprint to train a binary classifier[51], which achieved an accuracy of 83.8% on the test set (Section S7).

We utilized Shapley Additive Explanations (SHAP)[52] to quantify feature contributions across the fingerprint set. As shown in Figure 6a, the top 10 molecular substructures are ranked by mean absolute SHAP values, with CN([O])N(C)[O] emerging as the most globally important subunit. Furthermore, the bee swarm plot reveals that the presence of this feature consistently shifts the model predictions toward positive $\xi$ values, underscoring its crucial role in promoting SF. To evaluate the influence of the CN([O])N(C)[O] fragment, we compared the target properties between the groups containing this motif and those without it. The positive Cohen's $d$ values for $E_{S1}$ ($d = 0.83$) and $E_{T2}$ ($d = 0.59$) indicate that the fragment elevates these two excited-state energies, while slightly lowering $E_{T1}$ ($d = -0.18$) (Figure 6c)[53]. Consequently, the energy gaps between $E_{S1}$ and $E_{T1}$, as well as between $E_{T2}$ and $E_{T1}$, are both effectively widened. This trend is further corroborated by Figure 6d, which shows that the CN([O])N(C)[O] fragment occurs in approximately 42% of the analyzed molecules and exhibits the largest positive difference between its effects on $E_{S1}$ and $E_{T1}$. Taken together, these statistical and property analyses suggest that this fragment may serve as a key structural motif in the design of SF molecules.

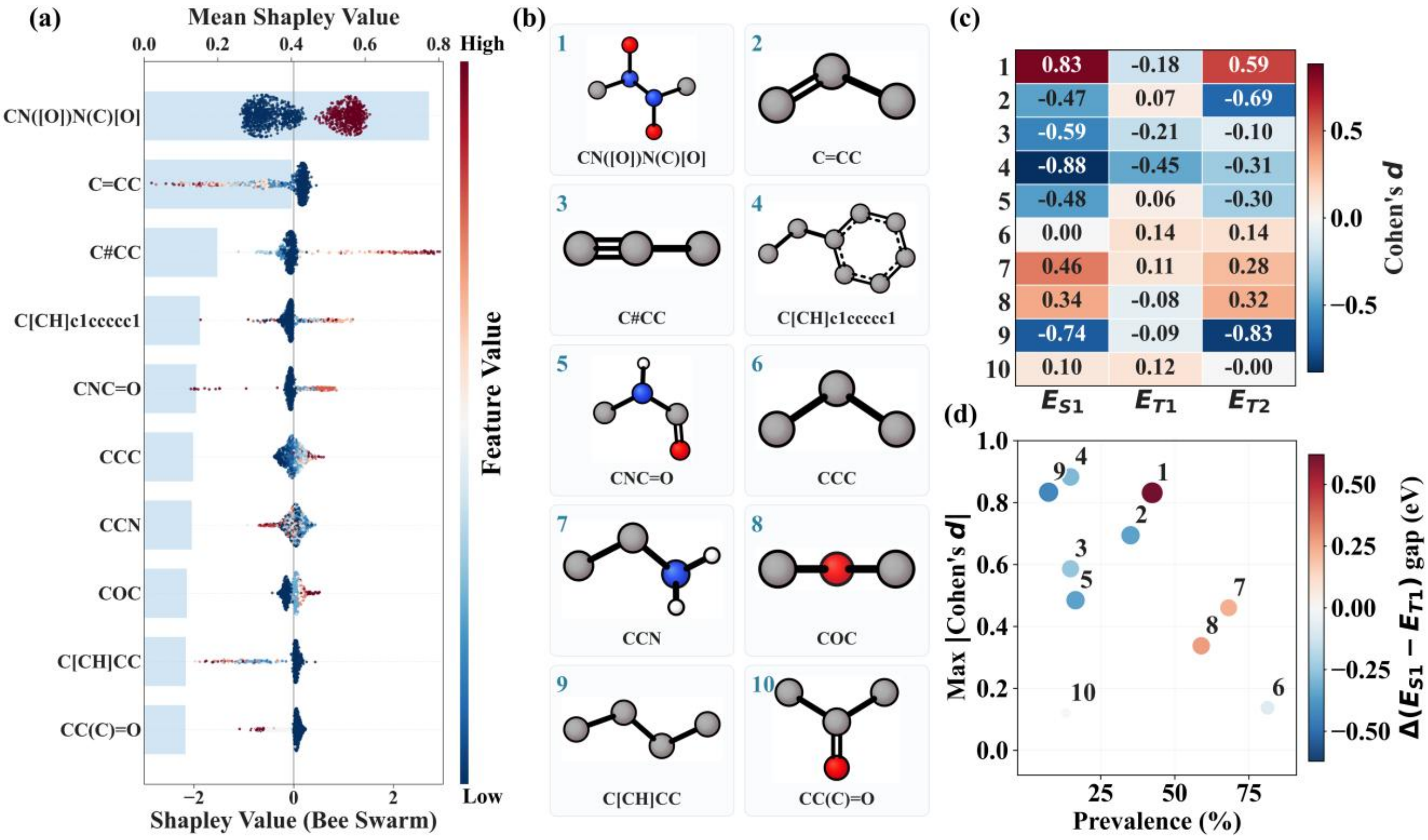

**Figure 6.** SHAP interpretation of the classifier for SF propensity. (a) Global SHAP summary of the ten most influential fingerprint fragments: the bar shows mean SHAP importance, and the bee swarm shows SHAP values. (b) Structures of the ten numbered fragments. (c) Fragment-resolved effects on $E_{S1}$, $E_{T1}$, and $E_{T2}$, shown as Cohen's $d$. (d) Comparison between feature importance (maximum absolute Cohen's $d$ among three energy descriptors) and fragment distribution. The bubble color encodes the difference in the average value of ($E_{S1}$−$E_{T1}$) between molecules containing and lacking the corresponding fragment.

Motivated by these results, a representative molecule incorporating the CN([O])N(C)[O] fragment and exhibiting favorable SF energetics ($\xi \geq 0$) was selected as a case study. As shown in Figure 7, the molecule can be deconstructed into three units: Cc1ccccc1 (F1), CN([O])N(C)[O] (F2), and c1ccc(C=N)cc1 (F3). Frontier molecular orbital analysis shows that the HOMO of the whole molecule is predominantly localized on the F2 fragment. This localization can be understood from the orbital energy alignment of the isolated fragments. The highest occupied orbital levels of F1 and F3 are substantially lower (−8.45 and −8.85 eV, respectively) than that of F2 (−7.48 eV), which is close to the HOMO energy of the whole molecule (−7.68 eV). In contrast, the LUMO distributes across both F2 and F3 due to orbital hybridization, driven by the small energy gap between their respective

LUMOs (0.87 eV for F2 vs. 0.60 eV for F3). Analysis of excited-state energy alignment further reveals that both $S_1$ and $T_1$ states predominantly originate from the F2 fragment. F2 exhibits a remarkably low triplet energy ($E_{T1}$ = 1.72 eV) that is preserved in the entire molecule, a feature likely attributed to the weak biradical character of F2 ($y_0$ = 0.18)[54, 55], which may also facilitate satisfying the $E_{T2} \geq 2E_{T1}$ condition. As a result, both the F2 fragment and the whole molecule satisfy the energetic criterion for SF. Furthermore, the electron-hole distributions of the three excited states indicate that $T_1$ and $T_2$ are predominantly localized on different fragments[56-58]. Such spatial separation is expected to reduce the wavefunction overlap between triplet excitons, thereby suppressing TTA. Importantly, incorporating the CN([O])N(C)[O] fragment does not guarantee favorable SF energetics. Although this fragment possesses a low triplet energy, the excited-state landscape of the full molecule is ultimately governed by the interplay between fragment energy levels and inter-fragment electronic coupling. As detailed in Section S8, if the excited states are dominated by other fragments instead of CN([O])N(C)[O], the system suffers from an unfavorable energy alignment, ultimately resulting in a negative SF Score.

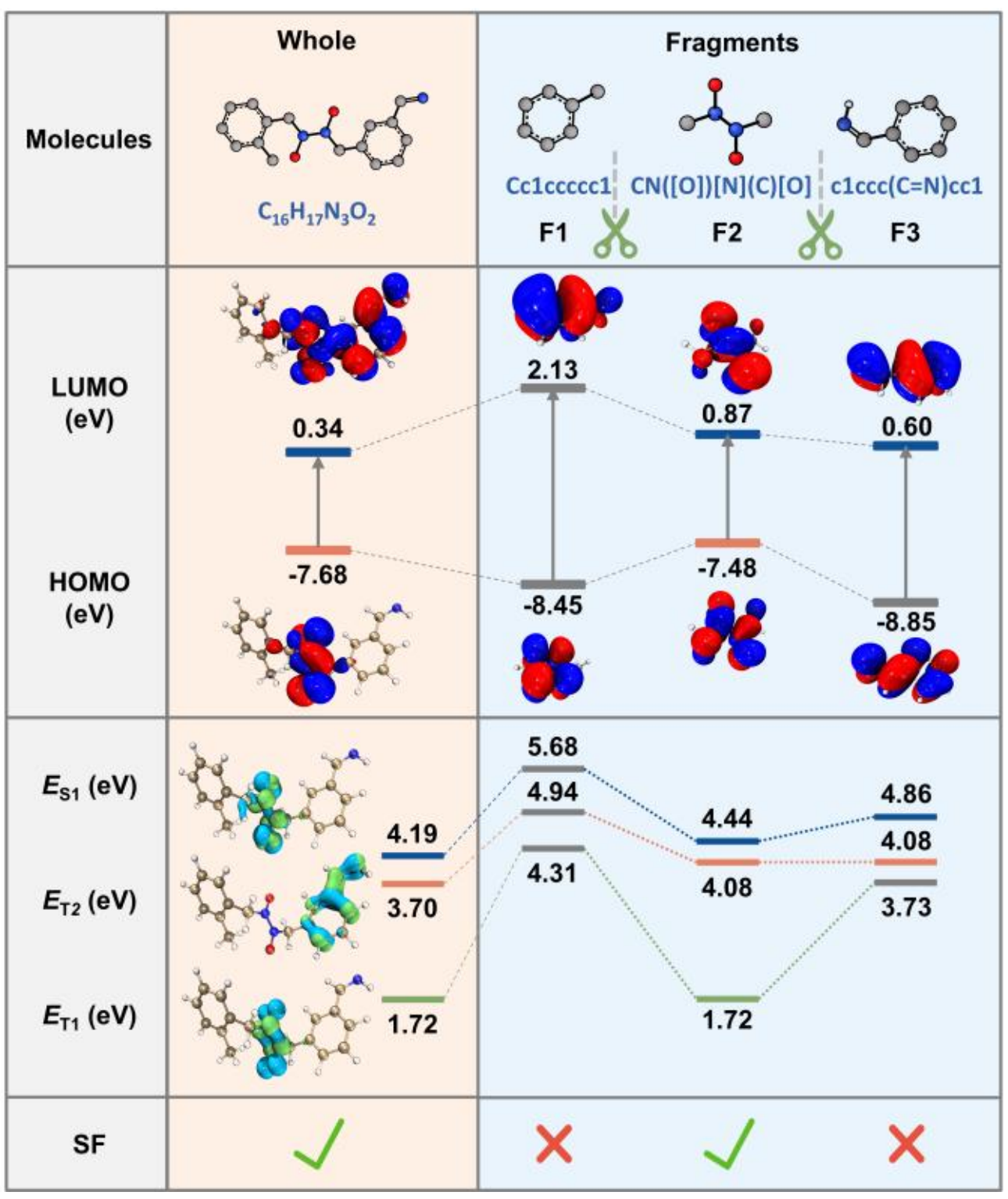

**Figure 7.** Fragment-resolved electronic-structure analysis of a representative molecule containing fingerprint feature 1, CN([O])N(C)[O]. Frontier molecular-orbital iso-surfaces (LUMO and HOMO) and electron-hole distributions for $S_1$, $T_2$, and $T_1$ are shown for the whole molecule and its F1, F2, and F3 fragments.

## CONCLUSION

We have developed a co-evolutionary generative-predictive framework that provides an efficient inverse-design route for SF molecules. By integrating molecular generation, excited-state prediction, quantum-chemical validation, and a unified SF scoring metric, the framework enables iterative optimization that goes beyond conventional forward screening. Screening over 100 million molecules and validating with multi-criteria filters produced over 280,000 TDDFT-confirmed SF candidates, greatly enriching the SF chemical space with structurally diverse and synthetically accessible molecules. SHAP-based fingerprint analysis reveals key substructures that favorably tune SF energetics, yielding interpretable design rules. More broadly, the modular architecture of this framework allows seamless adaptation of its structure generator, property predictors, and validation workflow to other excited-state and optoelectronic molecular design challenges, establishing a general and scalable paradigm for data-driven molecular discovery.

## ASSOCIATED CONTENT

### Data Availability Statement

The data that support the findings of this study are available at: https://doi.org/10.5281/zenodo.22718261. All codes are publicly available on the GitHub repository at: https://github.com/llf12138/SF_Gen-Pred. A package for SF molecule design, to which this work has contributed, is also available at: https://github.com/fuli-phy/SF_Mol_Designer.

### Supporting Information

The Supporting Information is available free of charge at XXX.

Details of models training and SMILES augmentation; definition of SF Score; synergistic generative-predictive framework; synthetic-accessibility assessment; construction of excited-state and singlet fission databases; binary classifier performance; and electronic structure analysis (PDF)

## AUTHOR INFORMATION

### Corresponding Authors

**Li Fu** − Guangdong Basic Research Center of Excellence for Structure and Fundamental Interactions of Matter, Guangdong Provincial Key Laboratory of Quantum Engineering and Quantum Materials,

School of Physics, South China Normal University, Guangzhou 510006, China; MOE Key Laboratory of Environmental Theoretical Chemistry, South China Normal University, Guangzhou 510006, China; E-mail: fuli@scnu.edu.cn

**Lingzhi Zhao** − Guangdong Provincial Engineering Technology Research Center of Low Carbon and Advanced Energy Materials, School of Electronic Science and Engineering (School of Microelectronics), South China Normal University, Foshan, 528225, China; E-mail: lzzhao@scnu.edu.cn

**Jijun Zhao** − Guangdong Basic Research Center of Excellence for Structure and Fundamental Interactions of Matter, Guangdong Provincial Key Laboratory of Quantum Engineering and Quantum Materials, School of Physics, South China Normal University, Guangzhou 510006, China; MOE Key Laboratory of Environmental Theoretical Chemistry, South China Normal University, Guangzhou 510006, China; E-mail: zhaojj@scnu.edu.cn

**Authors**

**Longfei Lv** − Guangdong Basic Research Center of Excellence for Structure and Fundamental Interactions of Matter, Guangdong Provincial Key Laboratory of Quantum Engineering and Quantum Materials, School of Physics, South China Normal University, Guangzhou 510006, China; MOE Key Laboratory of Environmental Theoretical Chemistry, South China Normal University, Guangzhou 510006, China

**Si Zhou** − Guangdong Basic Research Center of Excellence for Structure and Fundamental Interactions of Matter, Guangdong Provincial Key Laboratory of Quantum Engineering and Quantum Materials, School of Physics, South China Normal University, Guangzhou 510006, China; MOE Key Laboratory of Environmental Theoretical Chemistry, South China Normal University, Guangzhou 510006, China

**Notes**

The authors declare no competing financial interest.

## ACKNOWLEDGEMENTS

This work was supported by the National Natural Science Foundation of China (12534012, 12474168).

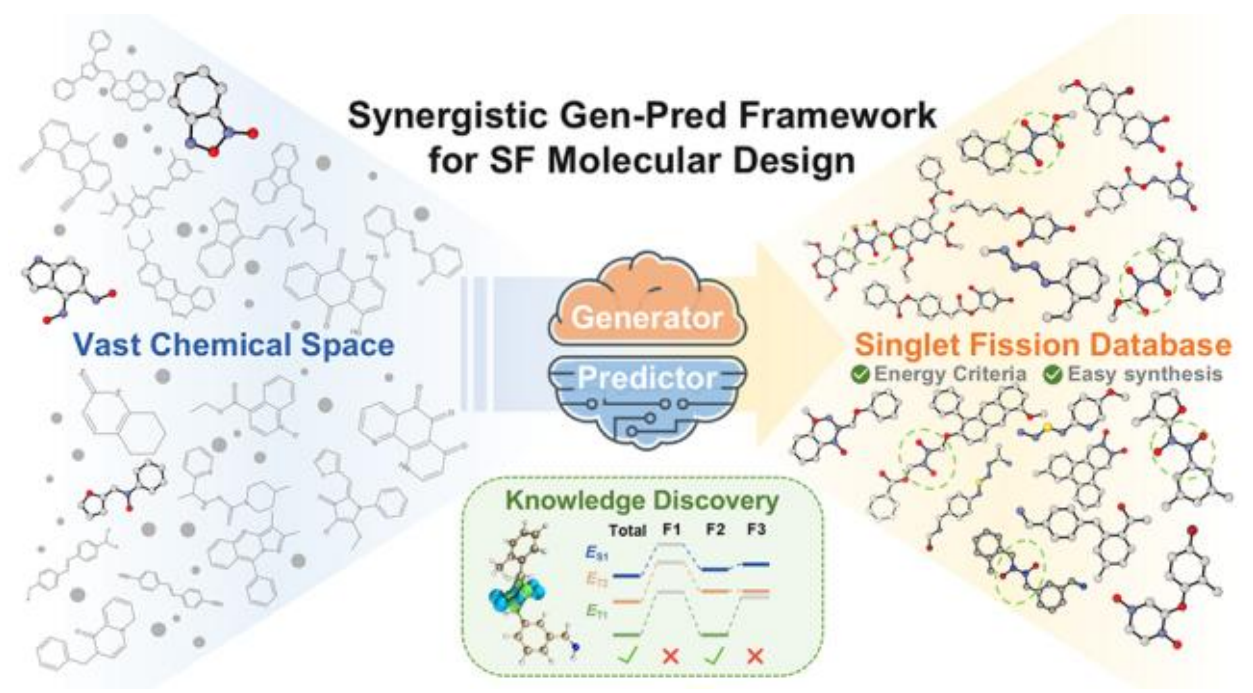

**For Table of Contents Only**